\documentclass[fleqn,usenatbib]{mnras}

\usepackage{newtxtext,newtxmath}
\usepackage[T1]{fontenc}

\DeclareRobustCommand{\VAN}[3]{#2}
\let\VANthebibliography\thebibliography
\def\thebibliography{\DeclareRobustCommand{\VAN}[3]{##3}\VANthebibliography}

\usepackage{graphicx}	
\usepackage{amsmath}	

\title[Hydrostatic and explosive $\alpha$-element ratios]{Hydrostatic and explosive $\alpha$-element chemical abundances of Milky Way globular clusters, halo substructures, and satellite galaxies}

\author[Horta $\&$ Ness]{
Danny Horta$^{1}$\thanks{E-mail: dhorta@roe.ac.uk} and
Melissa K. Ness$^{2}$
\\
$^{1}$Institute for Astronomy, University of Edinburgh, Royal Observatory, Blackford Hill, Edinburgh, EH9 3HJ, UK\\
$^{2}$Research School of Astronomy $\&$ Astrophysics, Australian National University, Canberra ACT 2611, Australia\\
}

\date{Accepted XXX. Received YYY; in original form ZZZ}

\pubyear{2025}

\begin{document}
\label{firstpage}
\pagerange{\pageref{firstpage}--\pageref{lastpage}}
\maketitle

\begin{abstract}
Stellar atmospheric element abundance ratios of stars retain information about their birth conditions, helping elucidate their origin and nature. 
In this letter, we analyse and contrast the hydrostatic and explosive $\alpha$-element abundance ratios, and the ratio of the two (the hex ratio), for a large sample of Galactic globular clusters (GCs), halo substructures, satellite galaxies, and the Milky Way high-/low-$\alpha$ discs using data from the \textsl{APOGEE} survey. Our results show that: $i$) Milky Way GCs and halo substructures appear to have qualitatively similar hex ratios across a broad range of [Fe/H], that are higher than that of dwarf satellite galaxies of similar [Fe/H]; $ii$) for all stellar populations studied, there is a trend in the hex ratio with [Fe/H]; $iii$) there is a weak trend in the hex ratio with respect to age for Galactic GCs, but not with initial or final GC mass; $iv$) there are no differences in the hex ratio between GCs formed \textit{in situ} versus those labelled as accreted. 

\end{abstract}




\section{Introduction}

The chemical abundance profile of stellar populations retain vital clues about their origin and formation environment \citep{Freeman2002}. Of the spectrum of elements, the $\alpha$ type (namely, those composed from multiples of the $\alpha$-particle, $^{4}_{2}\mathrm{He}$) are particularly interesting because they are created during the burning and explosion of massive stars that detonate as type II supernovae, one of the main drivers of the star formation and chemical enrichment of galaxies. Thus, the $\alpha$-elements are particularly useful for tracing the star formation histories of galaxies, especially when contrasted to heavier iron-peak elements produced predominantly during the explosion of type-Ia supernovae (i.e., the Tinsley-Wallerstein diagram, \citealp{Tinsley1979}). 

While the $\alpha$-elements are typically grouped into one category, not all $\alpha$-elements are produced in the same way. They can be broadly sub-categorised into ``hydrostatic'' and ``explosive'' $\alpha$-elements, depending on if they are formed during nuclear fusion in the core of massive stars  (e.g., O, Mg) or during the type II supernovae explosion (e.g., Si, Ca, Ti)\footnote{Some heavier $\alpha$-elements are also produced in white-dwarf supernovae \citep{Johnson2020}.}. Importantly, not all stars that detonate as type II supernovae produce the same amount of $\alpha$-elements. The amount of hydrostatic and explosive elements synthesised depends on the mass of the star; hydrostatic $\alpha$-elements are solely produced in the core of the most massive stars ($M_\star\approx15-30~M_{\odot}$), whereas explosive $\alpha$-elements are created from all core-collapse supernovae explosions ($M_\star>8~M_{\odot}$). Therefore, the ratio of the hydrostatic to explosive $\alpha$-elements, dubbed the ``hex'' ratio \citep{2013McWilliam,Carlin2018}, provides a way of inferring the production of $\alpha$-elements from the most massive stars to the total number of core-collapse supernovae.

As the hex ratio traces how many massive stars have contributed to the star formation of a system when compared to the total number of supernovae type II, comparing the hex ratio between different stellar populations provides a way to estimate the high-mass end of the initial mass function (IMF). For example, \citet{Carlin2018} examined the hydrostatic and explosive $\alpha$-elements in the Sagittarius dwarf spheroidal galaxy (Sgr dSph), finding that the system had lower overall hex abundances when compared to Milky Way stars of the same metallicity; from this the authors inferred that the Sgr dSph is consistent with having formed from an IMF that lacks the most massive stars (i.e., top-light or bottom-heavy IMF, see also \citealp{2013McWilliam}). Similarly, \citet{Blancato2019} examined the ratio of [Mg/Si] for Milky Way disc stars across [Fe/H] and age, finding that there were small variations that could not be reproduced with chemical evolution models that assume a fixed IMF \citep[see also][]{Horta2022}. 

In this letter, we set out to examine the abundance of hydrostatic and explosive $\alpha$-elements, as well as the ratio between the two (i.e., the hex ratio), in the Milky Way high-/low-$\alpha$ disc, globular clusters, halo substructures, and satellite galaxies. Our aim is to examine and contrast stellar populations formed at different ages and metallicities, and under a wide range of star formation histories, using the well measured $\alpha$-elements in red giant stars from the \textsl{APOGEE} survey.

The letter is organised as follows: in Section 2 we describe the data and the sample used; Section 3 presents the main results, including an examination of the hydrostatic and explosive $\alpha$-elements across different stellar populations, metallicity, age, and mass (the latter two only for Galactic globular clusters). We close by summarising and providing our concluding remarks in Section 4.

\section{Data}
We use the final spectroscopic data release (DR17) from the \textsl{APOGEE} survey (\citealp{Majewski2017, SDSSDR17}). \textsl{APOGEE} is a dual-helmisphere survey that collected near-infrared data using two high-resolution spectrographs \citep{Wilson2019} mounted on the $2.5$m telescope at Apache Point Observatory \citep{Gunn2006} in New Mexico, and the Du Pont telescope at Las Campanas Observatory \citep{BowenVaughan1973} in Chile. Stellar parameters and element abundances are determined using the \textsl{ASPCAP} pipeline \citep{Perez2015} based on the \textsl{FERRE} code \citep{Prieto2006}, using the line-lists from \citet{Cunha2017} and \citet{Smith2021}. The spectra were reduced by a customized pipeline \citep{Nidever2015}. All target selection criteria for \textsl{APOGEE} and \textsl{APOGEE-2} are described in \citet{Zasowski2013} and \citet{Zasowski2017}, respectively; details of \textsl{APOGEE}-north can be found in \citet{Beaton2021}, whereas information about \textsl{APOGEE}-south are contained in \citet{Santana2021}.

The ids for the stars belonging to the Galactic globular clusters (GCs) used in this work come from the GC value-added catalogue in \textsl{APOGEE} DR17 \citep[][]{Schiavon2024}, and include the following GCs: Djorg\,2, FSR\,1758, HP\,1, NGC\,0104, NGC\,0288, NGC\,0362, NGC\,1851, NGC\,1904, NGC\,2298, 
NGC\,2808, NGC\,3201, NGC\,4147, NGC\,4590, NGC\,5024, NGC\,5053, NGC\,5139, NGC\,5272, NGC\,5466, 
NGC\,5904, NGC\,6121, NGC\,6171, NGC\,6205, NGC\,6218, NGC\,6229, NGC\,6254, NGC\,6273, NGC\,6293, 
NGC\,6304, NGC\,6316, NGC\,6341, NGC\,6380, NGC\,6388, NGC\,6397, NGC\,6441, NGC\,6522, NGC\,6540, 
NGC\,6544, NGC\,6553, NGC\,6558, NGC\,6569, NGC\,6642, NGC\,6656, NGC\,6715, NGC\,6717, NGC\,6723, 
NGC\,6752, NGC\,6760, NGC\,6809, NGC\,6838, NGC\,7078, NGC\,7089, Palomar\,5, Palomar\,6, Terzan\,2, Terzan\,4, Terzan\,9. We follow the recommendation of \citet[][]{Schiavon2024} and select likely GC star candidate members by enforcing \texttt{dPOS}$<1$ (stars withing one Jacobi radii), \texttt{dRV}$<2$ (stars within $2\sigma$ from the GC's line-of-sight velocity distribution), and \texttt{dPM}$<2$ (stars within $2\sigma$ from the GC's proper motion distribution).


Halo substructure star candidates are identified using the selection criteria from \citet[][]{Horta2021, Horta2023}, and include the following substructures: \textsl{Aleph, Arjuna, Gaia-Sausage/Enceladus, Helmi Streams, Heracles, Icarus, I'itoi, Nyx, Pontus, Sequoia} (identified using three different criteria), \textsl{Sagittarius, Thamnos}, and \textsl{Wukong.}

Satellite galaxies (except the LMC/SMC) are determined using the \textsl{APOGEE} \texttt{MEMBERSHIPFLAG} bitmasks (see \citet[][]{Mead2024} for more details). Conversely, the ids for LMC/SMC candidates are obtained using the selection criteria from \citet[][]{Hasselquist2021}. All together, the satellite galaxies studied include: \textsl{Bootes I, Carina, Draco, Fornax, LMC, Sculptor, Sextans, SMC}, and \textsl{Ursa Minor}.

Lastly, the high-/low-$\alpha$ discs are determined using the following cuts in the $\alpha$-Fe plane:


\begin{itemize}
    \item \textit{low-$\alpha$}: $(-0.8<\mathrm{[Fe/H]}<-0.4\wedge  \mathrm{[Mg/Fe]}<0.185)~\vee\\~(-0.4<\mathrm{[Fe/H]}<-0.05 \wedge \mathrm{[Mg/Fe]}<\mathrm{-0.167 \times [Fe/H]+0.12)}~\vee\\~(-0.05<\mathrm{[Fe/H]}<-0.6 \wedge \mathrm{[Mg/Fe]}<0.12)$;
    \item \textit{high-$\alpha$}: $-0.8<\mathrm{[Fe/H]}\wedge \mathrm{not~low-}\alpha$.
\end{itemize}

In addition, we also impose the following selection criteria to all GCs, halo substructures, and satellite galaxies to ensure reliable element abundance ratio measurements for red giant stars: SNR $>50$; $\log~g<3.6$; $3,000 < T_{\mathrm{eff}}<5,500$ K.

\subsection{Accounting for light-element variations in globular clusters}

\begin{figure}
    \centering
    \includegraphics[width=\columnwidth]{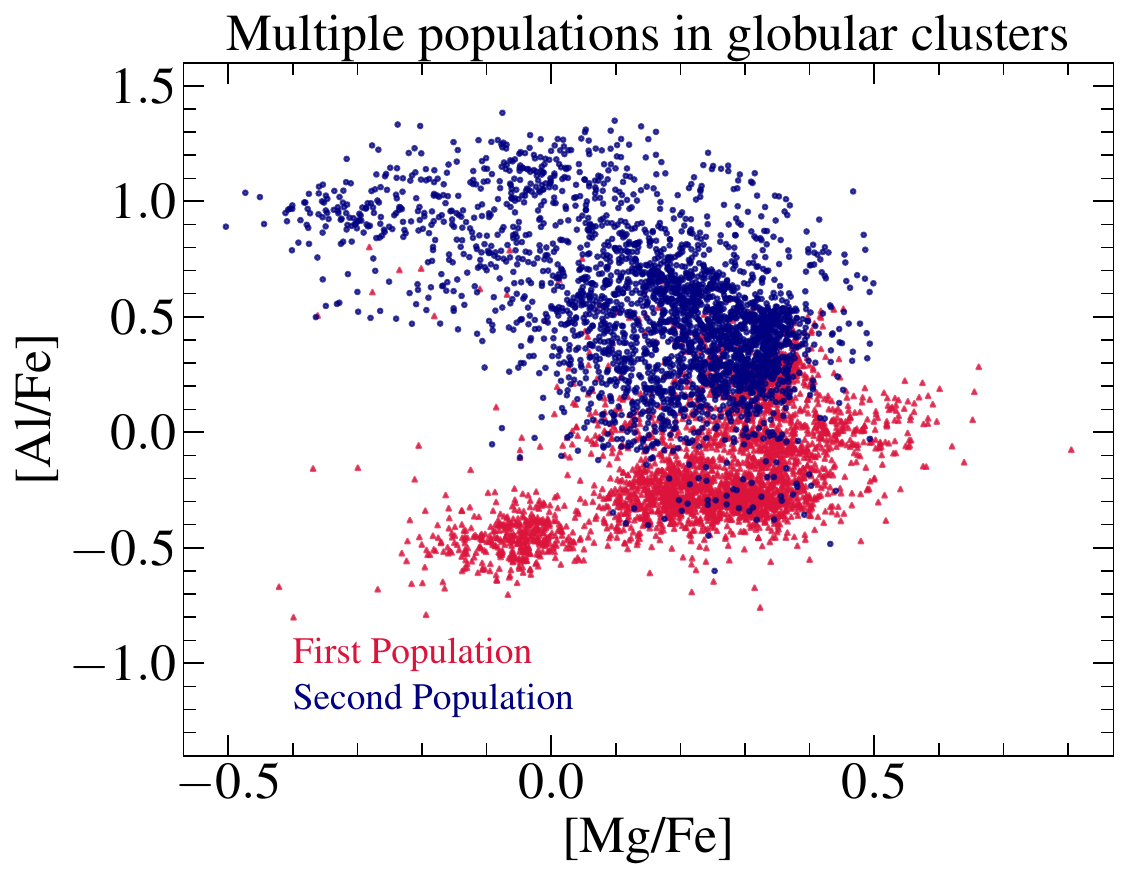}
    \caption{First-/Second-Population classification results from our $K$-means clustering procedure for stars in the GC value-added-catalogue \citep[][]{Schiavon2024} in the [Mg/Fe]-[Al/Fe] plane ([C/Fe]-[N/Fe] plane not shown). First population GC stars (red circles) do not suffer from light-element variations, and are thus used for studying the $\alpha$-element abundance ratios of Galactic GCs.}
    \label{fig_fpsp}
\end{figure}

GCs are known to host multiple-populations (MPs; \citealp{Bastian2018}). The second generation/population type (SP) are known to host light-element variations. Typically, SP stars are enriched in certain light elements whilst simultaneously being depleted in others, and this chemical profile isn't manifested in the first population (FP) type. For example, SP stars typically show enhanced levels of Al and depleted in Mg; the same effect happens for an enhancement in either N or Na and depletion in C and O, respectively \citep[e.g.,][]{Schiavon2017,Meszaros2020}. Thus, in order to mitigate any possible bias in our $\alpha$-element analysis, we must account for these chemical abundance anomalies arising from SP stars. To do so, we employ a $K$-means (unsupervised) clustering algorithm, in order to determine the FP GC stars that we will use in our analysis. We run this clustering procedure using the \texttt{kmeans} package in \texttt{scikit-learn} \citep[][]{scikit-learn}, and employing the following set of abundances for all stars belonging to each GC: [Mg/Fe], [Al/Fe], [C/Fe], [N/Fe]. When running the code, we choose the following hyperparameters: \texttt{n-clusters}$=2$; \texttt{random-state}$=0$; \texttt{n-init}$=$auto. We find that using this set-up, we are able to successfully split FP from SP GC stars (see Fig~\ref{fig_fpsp}). We also tested using a support vector machine to assess if we could obtain a better classification, but found that the $K$-means clustering provided similar classification performance.

\section{Hydrostatic and explosive $\alpha$-elements and the ``hex'' ratio}

\begin{figure*}
    \centering
    \includegraphics[width=\textwidth]{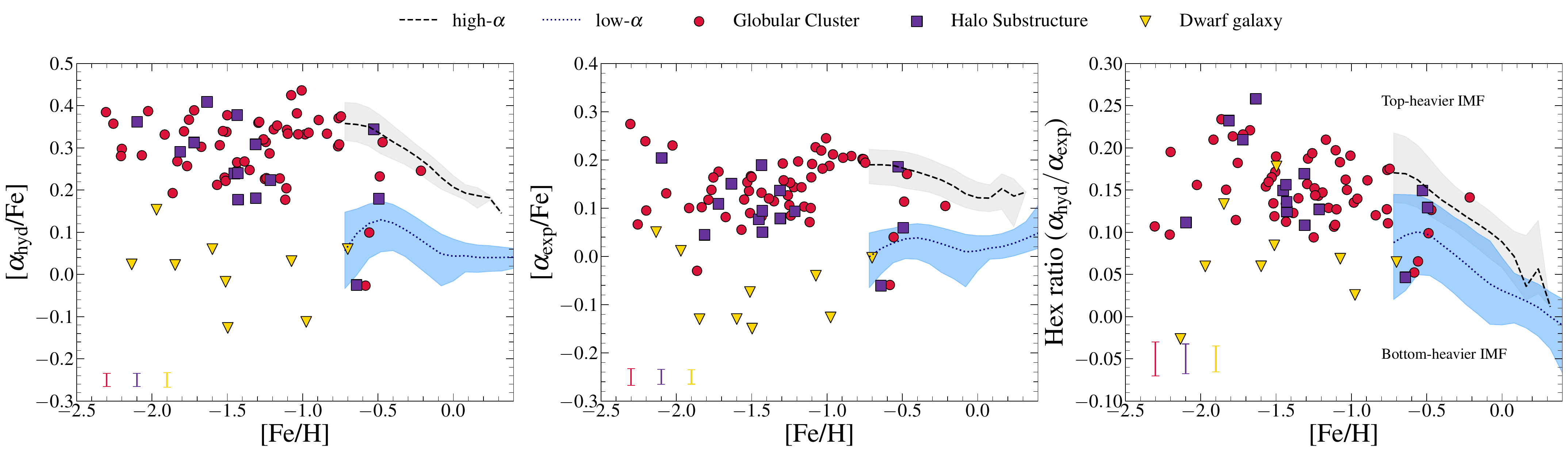}
    \caption{Median values of the hydrostatic $\alpha$-elements (namely, ([Mg/Fe]$+$[O/Fe])/2) on the left, the explosive $\alpha$-elements (namely, ([Si/Fe]$+$[Ca/Fe]$+$[Ti/Fe])/3) in the middle, and the hex ratio ([$\alpha_{\mathrm{hyd}}$/Fe] $-$ [$\alpha_{\mathrm{exp}}$/Fe]) on the right as a function of median [Fe/H] for (first population) globular cluster stars (red circles), halo substructures (purple squares), and Milky Way satellite galaxies (yellow upside-down triangles); the running median and [16,84]$^{\mathrm{th}}$ percentile range for the Galactic high-/low-$\alpha$ discs are also shown as dashed/dotted lines, respectively. Overall, GCs and halo substructures occupy a similar locus in all of these diagrams, which sits above the dwarf satellite galaxy sample. The difference in the hex ratio between the GCs and halo substructures with respect to the dwarf galaxies ---at fixed [Fe/H]--- is on the order of $\approx0.1$ dex. This implies that dwarf galaxies likely formed from a more top-light IMF when compared to Galactic GCs and halo substructures. The error bars show the average standard error for each stellar population type.}
    \label{fig_3panel}
\end{figure*}

Fig~\ref{fig_3panel} shows the median value of the hydrostatic $\alpha$-elements (namely, ([Mg/Fe]$+$[O/Fe])/2) on the left, the explosive $\alpha$-elements (namely, ([Si/Fe]$+$[Ca/Fe]$+$[Ti/Fe])/3) in the middle, and the hex ratio ([$\alpha_{\mathrm{hyd}}$/Fe] $-$ [$\alpha_{\mathrm{exp}}$/Fe]) on the right as a function of median [Fe/H] for (first population) GC stars (red circles), halo substructures (purple squares), Milky Way satellite galaxies (yellow upside-down triangles); we also show the running median and [16,84]$^{\mathrm{th}}$ percentile range for the Galactic high-/low-$\alpha$ discs as dashed/dotted (gray/blue) lines, respectively

Overall, we find that GCs and halo substructures occupy a similar locus in all of these diagrams, which sits above the dwarf satellite galaxy sample. In [$\alpha_{\mathrm{hyd}}$/Fe] and [$\alpha_{\mathrm{exp}}$/Fe], the difference between GCs and halo substructures with respect to dwarf galaxies is of $\approx0.3$ dex; conversely, the difference in the hex ratio between these populations is smaller, on the order of $\approx0.1$ dex. Furthermore, we find that the GCs and halo populations show a decreasing trend in [$\alpha_{\mathrm{hyd}}$/Fe] and the hex ratio with increasing [Fe/H], but approximately a flat profile in [$\alpha_{\mathrm{exp}}$/Fe]. This is also the case for dwarf galaxies, albeit there being an offset with the GCs/halo substructures.

Interestingly, we find that there is approximately a $\approx0.1$ dex difference between the average hex values of the high-/low-$\alpha$ discs (hex$_{\mathrm{high-\alpha}}\approx 0.15$ and hex$_{\mathrm{low-\alpha}}\approx 0.05$). As seen with other stellar populations examined, the hex ratio for the high-/low-$\alpha$ discs decreases with increasing [Fe/H]. This trend is likely caused by the decreasing ratio of [$\alpha_{\mathrm{hyd}}$/Fe], but approximately constant [$\alpha_{\mathrm{exp}}$/Fe], with increasing [Fe/H].

\subsection{Hex ratio across age and mass --- globular clusters}
\begin{figure}
    \centering
    \includegraphics[width=\columnwidth]{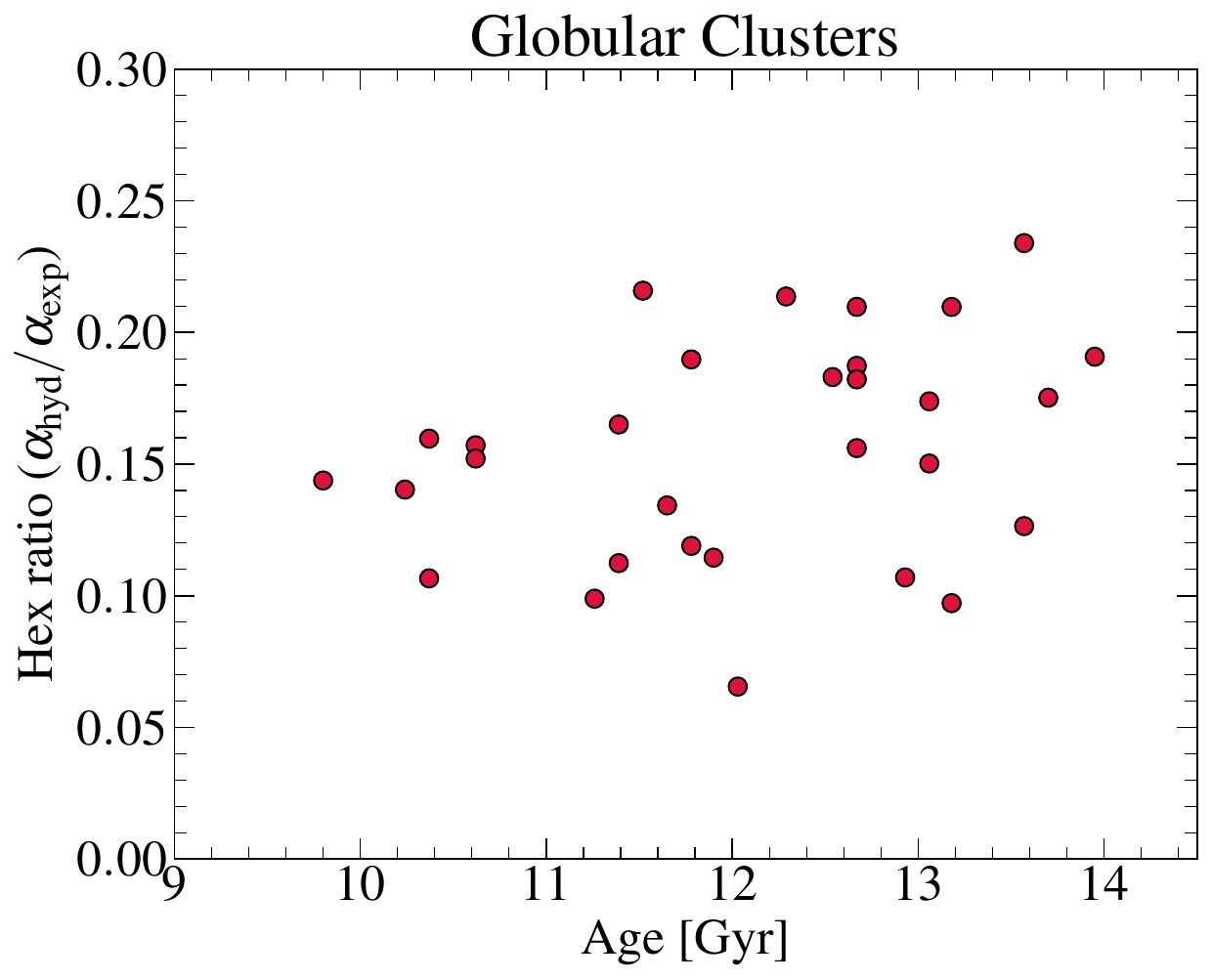}
    \caption{Median hex ratio as a function of age for Galactic GCs. Overall, there is a weak trend in hex ratio with age. However, we note there is also a significant scatter around this trend.}
    \label{fig_age}
\end{figure}

Fig~\ref{fig_age} shows the median hex ratio now as a function of age for the Galactic GC sample. The ages are taken from the compilation of \citet[][]{Forbes_2010}. Our results show that there appears to be a weak decreasing trend of hex with decreasing age\footnote{We calculate a slope of $\approx0.012$ dex with increasing age and an intercept of $\approx0.016$ dex.}. However, there is a significant scatter around this relation. 

We also investigated the relation between the hex ratio and GC mass ---both present day and initial mass--- using the values from \citet{Baumgardt2023} (not shown). We found there was no correlation between average hex ratio and GC mass.

\subsection{Hex ratio of accreted vs \textit{in situ} globular clusters}
\begin{figure}
    \centering
    \includegraphics[width=\columnwidth]{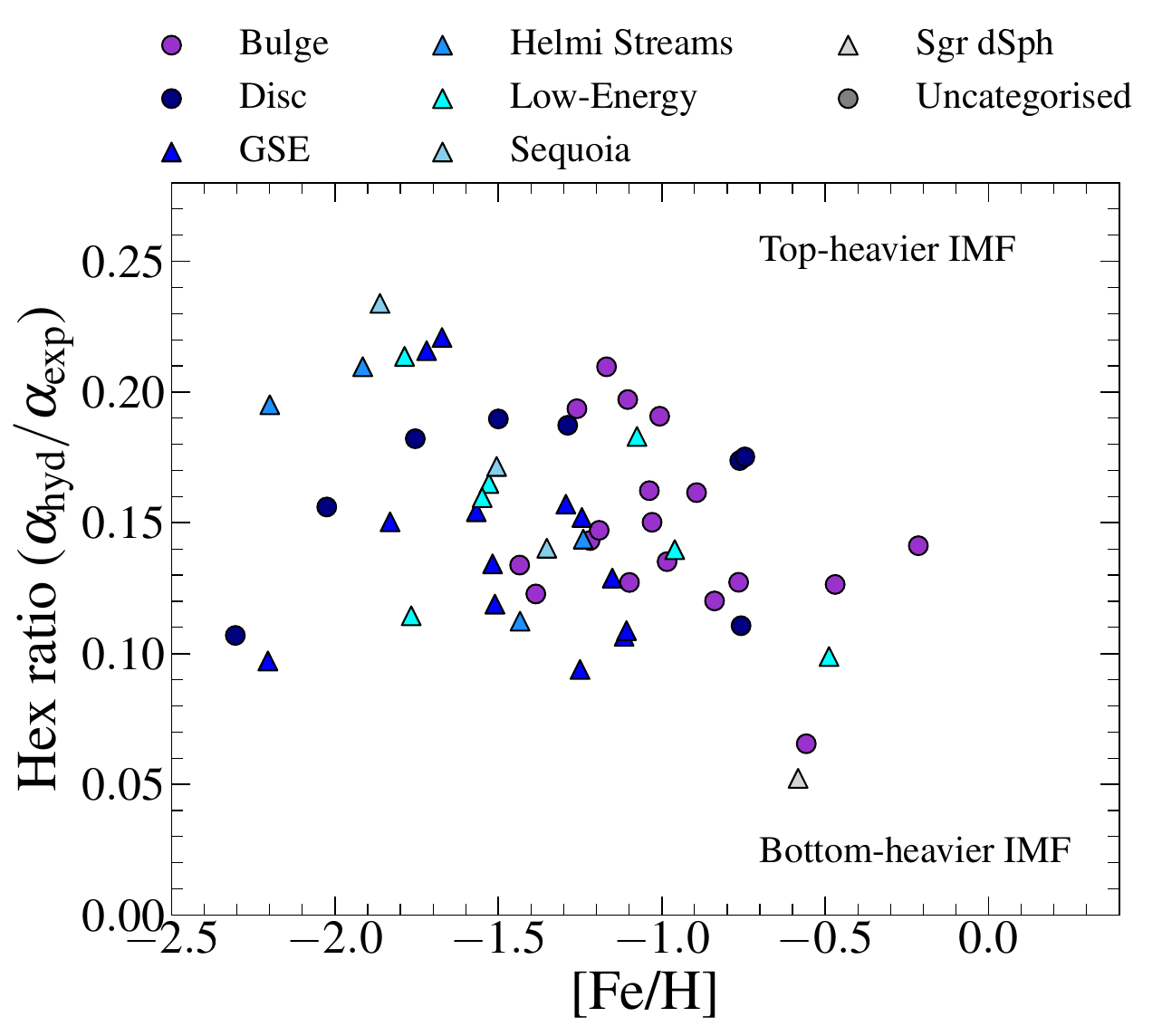}
    \caption{Average hex ratio as a function of median [Fe/H] for Galactic GCs, now divided into different origin subgroups based on their orbital properties and the age-[Fe/H] relation of GCs \citep[see][for details]{Massari_2019}. There is no evidence for different hex ratios among the different GC subgroups.}
    \label{fig_acc_insitu}
\end{figure}

Fig~\ref{fig_acc_insitu} shows again the median hex ratio as a function of median [Fe/H] for Galactic GCs, but now each GC is categorised into an accreted (triangles) or \textit{in situ} (circles) origin based on the associations from \citet[][]{Massari_2019}. We find that there is no dependence on origin between the different GC subgroups. All GCs follow the same decreasing trend in hex with increasing [Fe/H]. There are two clear outliers however, NGC\,7078 which is a [Fe/H]-poor ([Fe/H] $\approx-2.3$) disc GC, and NGC\,6341 ([Fe/H] $\approx-2.2$) that is a GSE GC. Both of these GCs present a low hex value for their metallicity (hex $\approx0.1$).

\section{Summary $\&$ conclusions}

At constant metallicity, a higher hex ratio value implies that a stellar population has undergone a star formation history that has endured a higher amount of core-collapse supernovae from the most massive stars; this is analogous to saying that a system was formed from a top-heavier IMF. Conversely, a lower hex ratio implies a smaller contribution from the most massive stars, and therefore, a more top-light IMF. In this work, we have set out to examine the hydrostatic to explosive $\alpha$-element abundance ratio in a range of stellar populations (namely, GCs, halo substructures, satellite galaxies, and the Galactic discs), all observed with the \textsl{APOGEE} survey. 

Fig~\ref{fig_3panel} reveals that overall Galactic GCs and halo substructures have similar hex ratios across a wide range in [Fe/H]. This hex ratio is more enhanced in these systems when compared to Milky Way satellite galaxies that ---at fixed [Fe/H]--- have on average a hex ratio $\approx0.1$ dex lower. This lower overall hex abundance in satellite galaxies is not due to a lower amount of hydrostatic $\alpha$-elements alone, as Milky Way satellites appear to also be depleted in explosive $\alpha$-elements when compared to Galactic GCs and halo substructures (see the left and middle panels of Fig~\ref{fig_3panel}). At face value, this qualitative examination implies that the Milky Way satellite galaxies studied underwent a star formation history that is consistent with them being formed under a more top-light IMF when compared to Milky Way halo substructures and GCs.

Extra-Galactic studies modelling the luminosity of galaxies ranging in mass and environment have found that those with higher star formation rates tend to have formed from a more top-heavy IMF \citep{Hoversten2008, Meurer2009, Lee2009, Gunawardhana2011, Cappellari2012, Weidner2013, Li2017,Demasi2018}; follow up studies also suggest that the IMF possibly varies with metallicity or time \citep{Li2023,Martin2024, Bate2025}. Assuming that the cause for the variation in the hex ratio at constant metallicity between the GC/halo substructures and the living satellites is due to variations in their star formation histories, and thus the high-mass end of the IMF, our results support these extra-Galactic studies.


Interestingly, when examining the distribution of the mean hex ratio as a function of average [Fe/H] for all stellar populations examined, we have found that GCs, halo substructures, and the Milky Way high-/low-$\alpha$ disc follow a relation in which the hex ratio decreases with increasing [Fe/H] (see also \citealp{Blancato2019}). The best-fit line to this relationship has a slope of $m = -0.047$ and an intercept of $b = 0.089$. Along the same lines, the Milky Way satellite sample also follows a linear trend, albeit different to that of the GCs/halo substructures. The Milky Way satellites' best-fit line has a slope of $m_{\mathrm{sat}} = -0.048$ and an intercept of $b_{\mathrm{sat}} = -0.050$.

We reason that the origin of the hex ratio trend with respect to [Fe/H] is due to the contribution of explosive $\alpha$-elements from SNIa explosions \citep{Johnson2020}; at higher [Fe/H] there have likely been more white dwarf supernovae detonations (that have a delayed onset of $\approx100$ Myr when compared to SNII). This leads to a higher contribution of both Fe and explosive $\alpha$-elements (e.g., Si, Ca, and Ti). As a result, [$\alpha_{\mathrm{hyd}}$/Fe] decreases with increasing [Fe/H] but [$\alpha_{\mathrm{exp}}$/Fe] remains approximately constant with increasing [Fe/H] (see the middle panel of Fig~\ref{fig_3panel}), yielding a decreasing hex ratio overall. However, in order to test this hypothesis, a full characterisation of the star formation histories for each individual system across [Fe/H] is needed.

Interestingly, we also find that there are some outliers in each sample studied. For example, the Sagittarius dSph (Sgr) and its core (M54) have a hex ratio that more closely follows the satellite galaxy trend (see also \citealp{2013McWilliam,Carlin2018}). This is perhaps not that surprising, given that Sgr is a satellite galaxy that has recently been accreted by the Milky Way. In contrast, the Sextans dwarf galaxy has a hex ratio that places it in the middle of the GC/halo substructure distribution in the [Fe/H]-poor end (see Fig~\ref{fig_appendix}). This result implies that Sextans likely formed from a more top-heavy IMF when compared to its satellite galaxy counterparts with approximately the same [Fe/H]. We also find that several halo substructures (Pontus, Sequoia, and Wukong) have an almost identical hex ratio (and $\alpha$-Fe abundance) to GES, in concordance with the hypothesis from \citet[][]{Horta2023} that these are likely different parts of the GES debris. The same insight applies to Aleph and Nyx, which show $\alpha$-abundances and a hex ratio that is very similar to the high-$\alpha$ disc. In addition, there is one dwarf galaxy (\textsl{Bootes I}) that has a hex ratio of $\approx0$ at a low [Fe/H] $=-2.15$. This implies that this dwarf galaxy could have formed from a significantly more top-light IMF, although see \citet[][]{Filion_2022}. There are also two GCs (NGC\,6341 and NGC\,7078) that show a lower hex ratio ($\approx0.1$) for their low metallicity ([Fe/H] $\approx-2.3$). 

In addition to examining the distribution of hex ratios with respect to [Fe/H], in Fig~\ref{fig_age}, we have  explored how the hex ratio varies as a function of age and mass for Galactic GCs. Overall, we find evidence for a weak trend of an increasing hex value with increasing age, irrespective of [Fe/H] ($m_{\mathrm{age}} = 0.012$ and $b_{\mathrm{age}}=0.016$). We also investigated how the hex ratio varies with GC (initial and final) mass, but (surprisingly) found no correlation. This (weak) trend with age may possibly highlight a small variation in the high-mass end of the IMF with GC birth time. However, the scatter around this relation is significant, so we are unable to draw any strong conclusions from the data. It would be interesting to follow up this line of research with bigger samples, and including other field stellar populations with accurate age/mass measurements. 

Lastly, in Fig~\ref{fig_acc_insitu} we also compared the hex ratio with respect to [Fe/H] for Galactic GCs that are deemed to be formed \textit{in situ} or have been accreted during galaxy mergers. Overall, we found that there was no difference between the hex ratio of different GC subgroups, despite their origin. Regardless, we argue that it may be beneficial to analyse the hydrostatic and explosive $\alpha$-elements independently, in addition to their orbital properties and ages, as these abundance ratios may help further elucidate the origin of Galactic GCs \citep[e.g.,][]{Horta2020, Geisler2021}.

In summary, contrasting the hydrostatic and explosive $\alpha$-element chemical compositions (namely, the hex ratio) of stellar populations provides a way of inferring the high-mass end of the IMF. In this work, we have found that Milky Way satellite galaxies have overall a lower hex ratio when compared to Galactic GCs and halo substructures at all [Fe/H]. It will be interesting to follow up this analysis by contrasting different heavier neutron-capture subgroups ($r/s$-process; e.g., \citealp{Monty2024,Henderson2025}), as well as other elemental species (e.g., \citealp{Hasselquist2021,Fernandes2023}) to gain further insight into the IMF and star formation histories from which these systems formed. Furthermore, if accurate stellar ages were available, it would also be interesting to quantify how the difference in the hex ratios between stellar populations manifests in terms of different IMFs; a possible avenue for exploring this could involve flexible Galactic chemical evolution modelling. 

\section*{Acknowledgements}
The authors thank the anonymous referee for their helpful report. DH would like to thank Jennifer Mead and Andrew Mason for help using the bitmasks in \textsl{APOGEE} to identify dwarf galaxy star candidates, and Andrea Kunder for helpful discussions. DH would also like to thank Sue, Alex, and Debra for their support. This work was supported by the UKRI Science and Technology Facilities Council under project 101148371 as a Marie Curie Research Fellowship. 

\section*{Data Availability}
\textsl{Data:} All \textsl{APOGEE} data used in this study is publicly available and can be downloaded directly from \url{https://www.sdss4.org/dr17/}.

\textsl{Software:}
    Matplotlib \citep{Hunter:2007},
    NumPy \citep{numpy},
    scikit-learn \citep{scikit-learn}.



\bibliographystyle{mnras}
\bibliography{refs} 

\appendix
\section{Hex ratio of individual globular clusters, halo substructures, and satellite galaxies}
\begin{figure*}
    \centering
    \includegraphics[width=\textwidth]{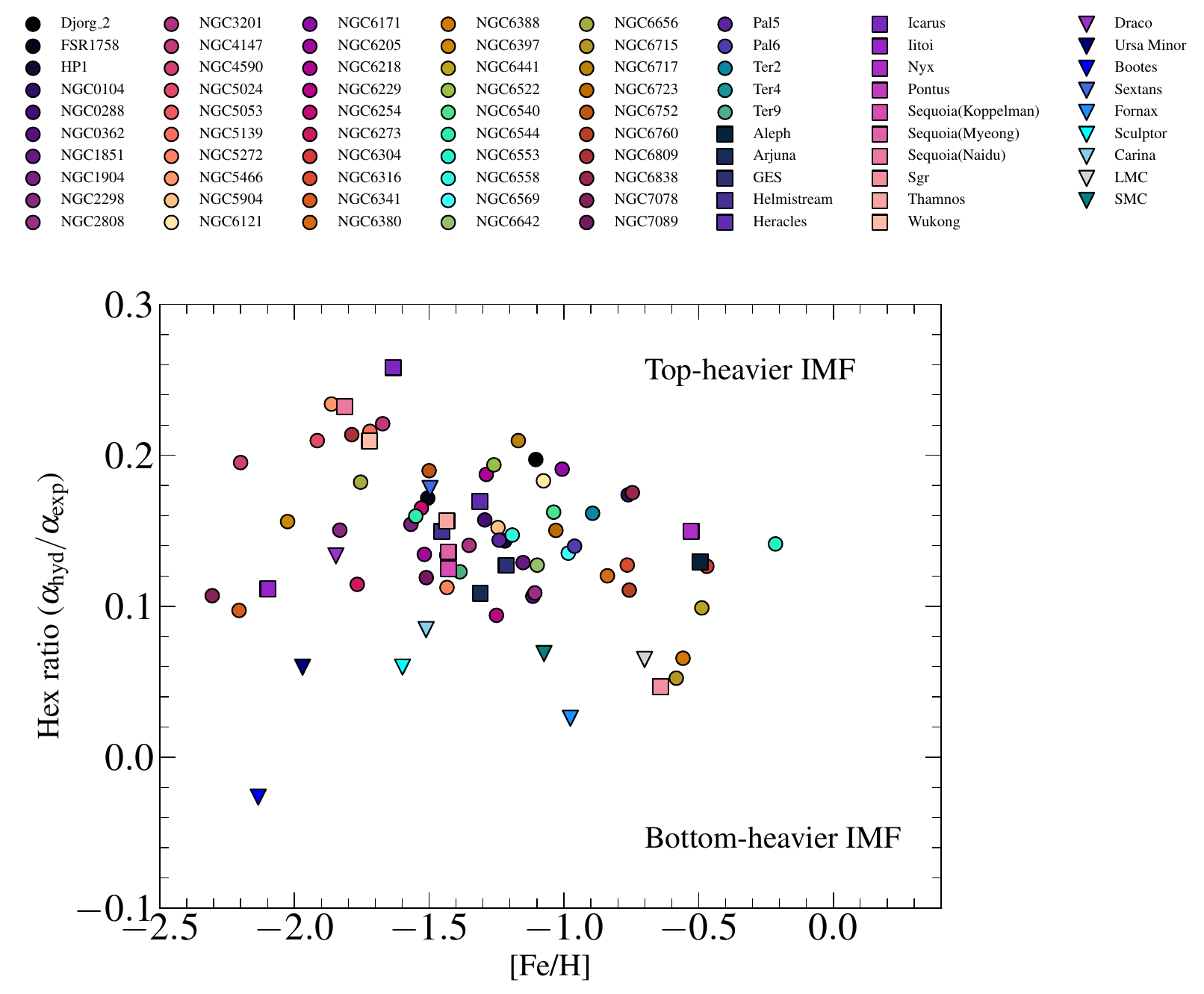}
    \caption{Average hex ratio as a function of median [Fe/H] for all samples studied in this work.}
    \label{fig_appendix}
\end{figure*}
\bsp	
\label{lastpage}
\end{document}